\documentclass[useAMS,usenatbib]{mn2e}
\usepackage{graphicx,amsmath,multirow,amssymb}
\usepackage{natbib}
\newcommand{\comment}[1]{}

\title[AGB modelling: the challenging interplay
between mass loss and molecular opacities]
{Asymptotic Giant Branch stars at low metallicity:\\
the challenging interplay between mass loss
and molecular opacities}
\author[P. Ventura and P. Marigo]{P. Ventura$^{1}$\thanks{E-mail:
ventura@oa-roma.inaf.it (AVR)} and P. Marigo$^{2}$\\
$^{1}$INAF-Osservatorio Astronomico di Roma, Via Frascati 33, Monte Porzio Catone 00040, Italia\\
$^{2}$ Department of Astronomy, University of Padova,
        Vicolo dell'Osservatorio 3, I-35122 Padova, Italy}
\begin{document}

\date{Accepted 1988 December 15. Received 1988 December 14; in original form 1988 October 11}

\pagerange{\pageref{firstpage}--\pageref{lastpage}} \pubyear{2002}

\maketitle

\label{firstpage}

\begin{abstract}
We investigate the main physical properties of low-metallicity Asymptotic Giant
Branch stars, with the aim of quantifying the uncertainties that presently
affect the predicted chemical yields of these stars,
associated to mass loss and description of molecular opacities.
We find that above a threshold mass,
$M~\simeq 3.5~M_{\odot}$ for $Z=0.001$,
the results are little dependent on the opacity treatment, as long as
hot-bottom burning prevents the surface C/O ratio from exceeding
unity; the yields of these massive AGB stars are expected
to be mostly determined by the efficiency of convection, with
a relatively mild dependence on the mass-loss description.
A much higher degree of uncertainty is associated to the yields of
less massive models, which critically depend on the adopted
molecular opacities.
An interval of masses exists, say $2.0-3.0~M_{\odot}$,
(the exact range depends on mass loss), in which HBB may be even
extinguished following the cooling produced by the opacity of C-bearing
molecules. The yields of these stars are the most uncertain,
the variation range being the largest (up to $\sim 2$~dex) for the nitrogen and
sodium yields. For very low-mass models, not experiencing
hot-bottom burning ($M\le 1.5~M_{\odot}$),
the description of mass loss and the
treatment of the convective boundaries are crucial
for the occurrence of the third dredge-up,
with sizable consequences on the CNO yields.
\end{abstract}


\section{Introduction}
The Asymptotic Giant Branch (AGB) is a common phase of the evolution
of stars with mass in the range 1--8M$_{\odot}$ \citep{busso, falk2}.
After core-helium burning, they experience a series of thermal pulses (TPs),
triggered by the ignition of helium in a thin layer below the CNO burning shell, 
under conditions of thermal instability. The strong mass loss that they experience
favours the ejection of the whole external mantle, with the formation
of a CO white dwarf.

Historically, the interest towards AGBs started with the discovery of
s-process enriched AGB stars \citep{merrill}, and with their identification as sites
of nucleosynthesis and s-process \citep{burbidge}, and was
futher stimulated by the discovery of carbon and lithium-rich stars in the 
Galaxy and the Magellanic Clouds (MCs) \citep{blanco,abia,sl1,sl2}: they 
were correctly interpreted as thermally pulsating stars \citep{sackmann,iben3}.
Extended investigations of the physical properties of AGB stars showed
the existence of two mechanisms able to alter the surface chemistry:
the Third Dredge-Up (TDU), i.e. the inwards penetration (in mass)
of the external envelope after each TP into regions earlier involved 
in $3\alpha$ nucleosynthesis \citep{iben1, fuji76};
and Hot Bottom Burning (HBB), when the base of the
convective envelope becomes sufficiently hot ($\geq 40\times 10^6$K)
to produce proton-capture nucleosynthesis \citep{scalo75, alvio}, and
enough energy to cause the breakdown of
the core mass-luminosity relation on the AGB \citep{blo2, booth92}.

In more recent years, we have been gathering the compelling evidence that
TP-AGB stars play a crucial  role in many properties of their host
systems, mainly owing to their intrinsic brightness
and distinctive spectral features.
For instance, the TP-AGB contribution to the total luminosity
of single-burst stellar populations reaches a maximum of about 40\% at
ages from 1 to 3 Gyr \citep{frogel90}, and accounts for most of the
bright-infrared objects in resolved galaxies, as clearly demonstrated
by DENIS, 2MASS, SAGE, S$^3$MC, and AKARI IRC data
\citep{cioni99, nik00,blum06,bol07,ita08}
for the Magellanic Clouds (MC).
Observational data of AGB stars is now available also for other galaxies
of the Local Group -- such as
Leo~I \citep{held10}, Leo~II \citep{gull08}, Sagittarius
\citep{gull07},
Phoenix \citep{menzies2}, Fornax \citep{patricia} --,
as well as for more distant dwarf galaxies
(dalcanton 2008)\footnote{HST prop. 11719}.

Chemical pollution from massive AGB stars is currently one of the most plausible hypothesis
to account for the chemical patterns observed in Globular Cluster (GC) stars
\citep{carretta}, where they may explain the existence of multiple populations
\citep{paolo3}, indicated
by the discovery of multiple main sequences in some GCs
\citep{piotto}.
The pioneering model
by \citet{annibale} outlines the capability of massive AGB stars to provide an
efficient pollution of the interstellar medium, and to stimulate the formation of
new stellar generations, with predicted  photometric and spectroscopic properties 
consistent with observations.

Moving from zero to high redshift, the light contribution from AGB stars becomes
significant in the optical, where galaxies
are dominated by intermediate-age stars \citep{bres94, bc03, mar05}.
This fact is crucial: it has been recently pointed out that
quantifying the weight of TP-AGB stars has a large impact on the
mass assembly in high redshift galaxies \citep{santini09}.

The prominent role played by AGB stars in so many astrophysical contexts
has increased the demand for detailed AGB modelling.
Since full computations of the TP-AGB phase
are extremely time-consuming, given the short time steps that must be
used during each TP, many synthetic models have been developed
and are presently
used, in which, based on existing full models,
the AGB properties are parametrized
as a function of stellar mass and metallicity
\citep{martin, paola03, bob2, paola04}.

Unfortunately full AGB modelling is dependent on the assumptions on many
physical ingredients that are not known from first principles, and must
be described by means of semi-empirical calibrations.
The treatment of convection is ascribed most of the
large differences among AGB models in the existing literature:
the efficiency of the convective instability, and the assumptions
concerning the extension of the mixed regions have a strong
impact on the physical and chemical evolution of massive AGB stars
\citep{falk,falk2,paolo4}.

Recent investigations have also shown that the main evolutionary
properties of AGB stars can be critically affected by the radiative
low-$T$ opacities adopted to model their outer mantles.  The use
of opacities that correctly account for the drastic changes in the
molecular chemistry of the gas, when
the C/O ratio passes from below to above unity, is mandatory,
otherwise the general expansion
of the outer layers triggered by the opacity increase is missed
\citep{paola02}.
Since one of the most significant  consequences of a correct computation 
of the molecular opacities is a sudden increase in mass loss, a careful study
of the effects driven by molecular opacities 
on the evolutionary properties and stellar yields  
requires to investigate the impact of mass loss, and its uncertainties.

In order to produce more reliable full AGB models,
that may also contribute a useful input to synthetic AGB models,
we have started a new project dedicated,
in a first phase, to understand and quantify the
main uncertainties associated to the various input physics.

The properties of
the most massive AGB stars, and the relevant role played by convection modelling,
was analysed by \citet{paolo4}. Moving to lower masses, \citet{paolo09}
have discussed the role played by molecular opacities.
In this work we concentrate on how the opacity  treatment is interfaced with
the description of mass loss, and which results in the literature
need a substantial revision. Future steps will be an extensive comparison
of model predictions with observations,
to provide the astrophysical community with more reliable yields.

The paper is organized as follows. Section~\ref{sec_uncert}
summarizes the uncertainties affecting AGB models.
The physical ingredients of the models presented
in the present investigation are detailed in Sect.~\ref{sec_input}.
Section~\ref{sec_mod} is
dedicated to the analysis of the physical properties of the models, and
how they depend on the choices concerning opacity and mass loss. 
The reliability of the yields, and the expected chemical patterns, 
are discussed in Sect.~\ref{sec_yields}. A comparison with the results
from other  investigations on the same topic is presented
in Sect.~\ref{sec_compar}.

\section{About the uncertainties affecting AGB modelling}
\label{sec_uncert}
The theoretical description of the AGB evolution is
extremely sensitive to the input physics used to calculate the models:
the treatment of the convective instability, the description
of mass loss, the extent of the possible extra-mixing region beyond the formal
convective/radiative boundary determined by the classic Schwartzschild
criterion, the adopted low-temperature opacities, and the nuclear
cross-sections of some p- and $\alpha-$capture reactions. An exhaustive
analysis can be found in \citet{falk2}.

\citet{paolo4}, following the analysis by \citet{blo2} and \citet{alvio},
showed that convection modelling plays an important role in determining
the evolution during the TP phase: the differences in the convection 
modelling have been recognized as the main cause responsible
for the discrepancies 
among the results obtained by different groups \citep{dh03,fenner,kl07}.

The assumption of some convective overshoot\footnote{By convective overshoot
we mean here the further distance travelled by convective eddies beyond
the border where buoyancy vanishes, fixed by the Schwarzschild
criterium} from the bottom of the envelope is closely associated to the occurrence of TDU after a TP that, in turn,
is confirmed by the discovery of many carbon stars in the Galaxy and in the
Magellanic Clouds (see e.g. \citet{batti, gro09}).
The analyses by \citet{gro93} and \citet{paola03}, based on synthetic 
TP-AGB models, indicate that the extent of the TDU required to reproduce 
the luminosity function of the MCs should be larger than predicted by the 
standard models, where TDU is found in the context of the Schwarzschild 
criterion.

With respect to mass loss along the AGB, several prescriptions, based
on either theoretical or empirical grounds, have been proposed in the
literature. Among them we recall: Reimers' classical law (1975)
was a fit of measured
mass-loss rates as a function of $M R/L$;  \citet{VW93} (hereinafter
VW93) was based on the observed correlation between mass-loss rates
and periods of pulsating AGB stars;
\citet{blo} proposed a modification to the Reimers' law
based on fitting models from \citet{bowen88};
\citet{bowen91} presented a prescription based on dynamical
models of pulsating oxygen-rich atmospheres including the formation of
dust; basing on those models \citet{bedijn88} derived a mass-loss
formalism as a function of basic stellar parameters, $M$, $R$, and $T_{\rm eff}$;
\citet{wachter02, wachter08}
proposed a formula based on dust-driven superwind models of C
stars; \citet{jacco} presented an empirical calibration based on
observations of dust-enshrouded red supergiants and oxygen-rich AGB
stars; \citet{schr05}  introduced a correction to the Reimers' formula
based on some physical arguments;  \citet{oscar2} (hereinafter S06)
formulated a revised
calibration of the mass loss-period relation for pulsating AGB stars.

In AGB models, an additional source of uncertainty is related to
the adopted low-T radiative opacities, which
quantify the removal of radiation energy as photons pass through the
H-rich mantle. The early predictions by \citet{paola02} were confirmed
by \citet{sergio}, who showed how the physical and chemical evolution
of low-mass, low-Z stars
are affected by the adoption of the correct opacity treatment. 

\section{Physical and chemical inputs}
\label{sec_input}
The models presented in this paper were calculated by means of the ATON code
for stellar evolution \citep{italo}, a full description of which can be
found in \citet{paolo1}.

\subsection{Convective regions}
Convection was modelled according to the Full Spectrum of Turbulence (hereinafter
FST) prescription by \citet{canuto}. Mixing of chemicals and nuclear burning were
coupled by means of a diffusive approach, following the scheme presented
in \citet{clout}; accordingly, convective overshoot was modelled by an 
exponential decay of convective velocities beyond the formal borders, with 
an e-folding decay of $l=\zeta H_P$. During the two main phases of core nuclear 
burning and in occurrence of the second dredge-up we used $\zeta=0.02$, in agreement 
with the calibration based on the width of the main sequences of open clusters
given in \citet{paolo1}. During the TP-AGB phase, given the uncertainties outlined
in the previous section, no overshoot was considered.

\subsection{Mass loss}
To investigate the sensitivity of the results obtained on the mass loss
prescription, our choice was to compare the findings obtained with two
different prescriptions, namely \citet{blo} and \citet{oscar2}.
In the former case it is assumed, based on hydrodynamical simulations
\citep{bowen88}, a steep increase of mass loss with luminosity as the 
star enters the AGB phase. This is simulated by multiplying the canonical 
Reimers' formula by a luminosity power (L$^{2.7}$). The free
parameter entering the Reimer's prescription is $\eta_R=0.02$,
according to the calibration of the luminosity function of
lithium-rich stars observed in the Magellanic Clouds given in
\citet{paolo2}. 

In \citet{oscar2} the mass-loss rate is made vary
with the pulsation period in the fundamental mode: 
compared to VW93 the empirical revision by S06 
(see their figure 5), based on a compilation of more recent data,  
predicts larger mass-loss rates for pulsation
periods in the range $100-300$ days, whilst lower rates are 
associated to periods larger than 500 days.

\begin{figure}
\caption{Evolution of the mass-loss rate ($\dot M$ in
$M_{\odot}$/yr) as a function of the surface luminosity for  models
with initial masses of
3M$_{\odot}$ calculated with
different prescriptions for mass loss and molecular opacity
(see Table~\ref{tab_mod} for
the meaning of the various symbols).
Each point in the plot marks the quiescent stage of the
luminosity maximum preceding the occurrence of a thermal pulse.
The  solid lines connect the four evolutionary sequences at the
stage when the total mass of the star has been reduced
to 2.9, 2.7, 2.5 and 2.0~M$_{\odot}$.}
\label{3msun}
\end{figure}

\subsection{Equation of state \& opacities}
For the equation of state (EOS) we adopted 
the latest version of the OPAL EOS (2005) where available,
superseded by the \citet{saumon} EOS in the partial ionization regime.
The EOS is extended to the high-density, high-temperature regime
according to the description given in \citet{blo3}.

The radiative opacities were calculated following the OPAL prescription,
according to \citet{iglesias}. The conductive opacities were taken from
Poteckin (2006)\footnote{See the WEB page
www.ioffe.rssi.ru/astro/conduct/}.

As to the low-temperature opacities, i.e.
$1\,500\,{\rm K} \le T \le 10\,000\,{\rm K}$,
we used the same large data set of tables as in \citet{paolo09}, which
were computed with the \AE SOPUS tool (Marigo \& Aringer 2009)\footnote{\AE SOPUS 
web-interface at http://stev.oapd.inaf.it/aesopus} to
consistently follow the significant changes in the CNO surface
abundances caused by the TDU and HBB during the TP-AGB evolution.

The opacity set was designed to account for the complex interplay
between convection and nucleosynthesis, so that variations in the C/O
ratio can be driven by positive/negative changes
in both carbon and oxygen abundances.
For instance, while an increment of C/O
is predicted due to the TDU, when HBB takes place the C/O
ratio is expected either to decrease, as long as C is
converted into N by the CN cycle, or even to increase if
O is efficiently burnt in favour of N by the ON cycle
(see Sect.~\ref{ssec_nucleos}).

Variations in CNO abundances affect the low-temperature opacities
essentially in two ways, i.e. i) for
$T\la 3\,000$ K by modifying the equilibrium molecular pattern depending
on the C/O ratio and, to a less extent, ii) for larger temperatures by
changing the contributions of the CNO atoms to both the continuum and
line opacity. A detailed discussion can be found in Marigo \& Aringer
(2009; see their section 4.2) and Ventura \& Marigo (2009).

\subsection{Nuclear reactions}
The nuclear network included in the code is described in details
in \citet{paolo4}. The cross-sections of the 64 reactions considered
are taken from the NACRE compilation \citep{angulo}, with the exception
of $^{14}$N(p,$\gamma$)$^{15}$O, taken from \citet{luna}, and the three
proton-capture reactions of the Ne-Na cycle, taken from \citet{hale1}
for the $^{22}$Ne(p,$\gamma$)$^{23}$Na reaction, and from
\citet{hale2} for the two p-captures by sodium nuclei.

\begin{figure*}
\caption{Left panel: Evolution of the core mass as a function of the
current  stellar mass for models
with initial masses 2.5, 3 and 3.5 $M_{\odot}$. Each point on the
tracks corresponds to quiescent CNO burning phase before the occurrence
of each TP. Refer to Table~\protect\ref{tab_mod}
for the meaning of symbols. Right panel:
The same as the left panel, but referring to the evolution of
3.5, 4.0 and 4.5 $M_{\odot}$ models.
}

\label{mcore}
\end{figure*}

\section{Evolutionary properties}
\label{sec_mod}
\subsection{An overview of the models}
The stellar models discussed here were followed from the pre-MS phase to
almost the complete ejection of the external envelope.
The initial chemical composition of the gas is assigned
a total metallicity (mass fraction) $Z=0.001$ and  a degree of
$\alpha-$enhancement $[\alpha/{\rm Fe}]=+0.4$, with the reference solar
mixture taken from \citet{gs98}.

To investigate how much the results
are affected by the
the interplay between the use of the opacities accounting
for the CNO variations  and the mass-loss description
we calculated 4 sets of evolutionary models, designated with
S06H, S06C, BH, and BC, which differ in the adopted prescriptions as
outlined in Table~\ref{tab_mod}.
Specifically, we consider two formalisms for the mass loss, i.e.
Straniero et al. (2006) and Bl\"ocker (1995), and two
treatments of the low-$T$ opacities, depending on whether
the underlying chemical mixture is kept fixed or accounts for changes
in the CNO abundances.
In order to better disentangle the effects of each
prescription, the models cover all four combinations of the two parameters,
opacity and mass loss.

\begin{table}
\caption{Input prescriptions of the TP-AGB models}
\label{tab_mod}
\begin{minipage}{0.5\textwidth}
\begin{tabular}{cccl}
\hline
\hline
\multirow{2}*{model} & \multirow{1}*{low-$T$ opacities:} &
\multirow{2}*{mass loss} &
\multicolumn{1}{r}{\multirow{2}*{symbol/line}\footnote{Symbols and line styles used in the plots}}\\
 & CNO variations  & & \\
\hline
S06H &  no & Straniero et al. (2006) & $\blacktriangle$/dash \\
S06C &  yes  & Straniero et al. (2006) & $\displaystyle\circ$/long-dash\\
BH &   no & Bl\"ocker (1995) & $\vartriangle$/solid \\
BC &   yes  & Bl\"ocker (1995) & $\scriptstyle\blacksquare$/dot\\
\hline
\end{tabular}
\vspace{-0.75\skip\footins}
\renewcommand{\footnoterule}{}
\end{minipage}
\end{table}


%





\begin{table*}
\caption{Relevant properties of AGB models}
\label{yields}
\begin{tabular}{cccccccccc}
\hline
\hline

M/M$_{\odot}$ & NTP & M$_c/$M$_{\odot}$  &  $\log(T_{\rm bce}^{\rm max})$ & Y & $[^{12}$C/Fe]  &  $[^{14}$N/Fe] & $[^{16}$O/Fe] & [Na/Fe] & R(CNO) \\
\hline
BH models  &&&&&&&&&\\
\hline
6.00  &  24  &  1.029 &  8.05  &  0.347 &  -0.707 &  1.325 & -0.369 & 0.248 &  0.929  \\
5.50  &  28  &  0.987 &  8.05  &  0.340 &  -0.686 &  1.350 & -0.468 & 0.283 &  0.946  \\
5.00  &  33  &  0.947 &  8.02  &  0.329 &  -0.505 &  1.458 & -0.410 & 0.402 &  1.205  \\
4.50  &  34  &  0.916 &  8.00  &  0.312 &  -0.233 &  1.669 & -0.112 & 0.674 &  2.012  \\
4.00  &  33  &  0.888 &  7.97  &  0.293 &   0.090 &  1.781 &  0.191 & 0.935 &  2.850  \\
3.50  &  28  &  0.857 &  7.94  &  0.270 &   0.259 &  1.902 &  0.464 & 1.147 &  4.074  \\
3.00  &  24  &  0.822 &  7.88  &  0.250 &   0.796 &  1.942 &  0.678 & 1.123 &  5.518  \\
2.50  &  22  &  0.746 &  7.49  &  0.257 &   1.805 &  0.519 &  0.985 & 0.416 & 11.204  \\
2.00  &  20  &  0.702 &  7.05  &  0.261 &   1.564 &  0.463 &  0.622 & 0.346 &  6.015  \\
\hline
BC models  &&&&&&&&&\\
\hline
6.00  &  25  &  1.028 &  8.05  &  0.348 &  -0.706 &  1.323 & -0.360 & 0.252 &  0.929  \\
5.50  &  30  &  0.983 &  8.04  &  0.335 &  -0.667 &  1.362 & -0.470 & 0.285 &  0.971  \\
5.00  &  33  &  0.946 &  8.02  &  0.329 &  -0.527 &  1.479 & -0.390 & 0.418 &  1.258  \\
4.50  &  35  &  0.916 &  8.00  &  0.311 &  -0.175 &  1.609 & -0.133 & 0.614 &  1.795  \\
4.00  &  32  &  0.886 &  7.97  &  0.292 &   0.121 &  1.758 &  0.190 & 0.912 &  2.747  \\
3.50  &  27  &  0.856 &  7.94  &  0.266 &   0.360 &  1.872 &  0.494 & 1.104 &  4.012  \\
3.00  &  22  &  0.821 &  7.63  &  0.250 &   1.226 &  1.336 &  0.687 & 0.397 &  4.477  \\
2.50  &  18  &  0.742 &  7.25  &  0.257 &   1.684 &  0.516 &  0.988 & 0.396 &  9.322  \\
2.00  &  16  &  0.677 &  6.88  &  0.265 &   1.710 &  0.481 &  0.897 & 0.354 &  9.065  \\
1.50  &  12  &  0.648 &  7.40  &  0.251 &   0.310 &  0.044 &  0.400 & 0.030 &  1.129  \\
\hline
S06H models &&&&&&&&& \\
\hline
6.00  &  73  &  1.044 &  8.07  &  0.355 &  -0.320 &  1.648 & -0.608 & 0.014 &  1.743  \\
4.50  &  50  &  0.924 &  8.01  &  0.320 &   0.298 &  2.023 &  0.001 & 0.648 &  4.377  \\
4.00  &  46  &  0.900 &  7.99  &  0.304 &   0.404 &  2.133 &  0.193 & 0.948 &  5.729  \\
3.50  &  36  &  0.883 &  7.97  &  0.270 &   0.360 &  2.216 &  0.373 & 1.184 &  6.997  \\
3.00  &  39  &  0.841 &  7.93  &  0.261 &   0.677 &  2.457 &  0.696 & 1.837 & 12.570  \\
2.50  &  52  &  0.806 &  7.89  &  0.270 &   0.739 &  2.677 &  0.983 & 2.436 & 21.029  \\
2.00  &  50  &  0.784 &  7.82  &  0.271 &   1.268 &  2.652 &  1.033 & 1.940 & 22.075  \\
1.50  &  28  &  0.731 &  7.03  &  0.260 &   2.033 &  0.156 &  0.948 & 0.385 & 16.285  \\
\hline
S06C models &&&&&&&&& \\
\hline
5.00  &  54  & 0.956  &  8.03  &  0.334 &   0.018 &  1.856 & -0.250 & 0.377 &  2.907  \\
4.50  &  47  & 0.921  &  8.01  &  0.317 &   0.083 &  1.973 & -0.059 & 0.585 &  3.825  \\
4.00  &  40  & 0.890  &  7.98  &  0.295 &   0.389 &  2.109 &  0.283 & 0.997 &  5.580  \\
3.50  &  21  & 0.846  &  7.95  &  0.270 &   0.251 &  2.202 &  0.490 & 1.310 &  6.998  \\
3.00  &  31  & 0.830  &  7.92  &  0.251 &   0.675 &  2.270 &  0.747 & 1.518 &  9.168  \\
2.50  &  23  & 0.758  &  7.51  &  0.253 &   1.833 &  0.510 &  1.113 & 0.428 & 12.840  \\
2.00  &  21  & 0.702  &  7.03  &  0.260 &   1.534 &  0.476 &  0.621 & 0.338 &  5.714  \\
1.70  &  18  & 0.685  &  6.88  &  0.253 &   1.573 &  0.053 &  0.578 & 0.059 &  5.909  \\
1.50  &  15  & 0.675  &  6.76  &  0.253 &   1.388 &  0.107 &  0.469 & 0.082 &  4.040  \\
1.20  &  14  & 0.660  &  6.57  &  0.253 &   1.141 &  0.056 &  0.412 & 0.056 &  2.608  \\
1.00  &  13  & 0.671  &  6.37  &  0.259 &   2.131 &  0.073 &  1.311 & 0.144 & 23.516  \\
\hline
\end{tabular}
\end{table*}

The resulting physical and chemical properties of the TP-AGB models
described above are presented in Table~\ref{yields}. For each  stellar
mass we show the number of thermal pulses experienced by the star,
the final core-mass, the maximum temperature reached at the bottom
of the external envelope, plus further information concerning the
average content of the ejecta, namely the helium mass fraction, and
the C, N, O  and Na enhancement/depletion factors, in terms of the
quantities [X/Fe], where [X/Fe]=$\log$(X/Fe)-$\log$(X/Fe)$_{\odot}$.
The last column shows the ratio between the average C+N+O abundance in
the ejecta and the initial value, which is assumed to represent the
chemical mixture at the epoch of the star's formation.

\subsection{The interplay between mass loss and molecular opacities}
We can appreciate the qualitative effects of the different  descriptions
of mass loss and molecular opacities from Fig.~\ref{3msun}, where
we show the evolution of a 3M$_{\odot}$ model calculated according to
the prescriptions listed in Table~\ref{tab_mod},
to which we refer for the meaning of the
various symbols. Core H- and He-burning phases are not included in
this plot, that starts from the beginning of the TP-AGB phase. Each
point marks the quiescent stage of pre-TP luminosity maximum.
The solid
lines are
iso-mass locii, and connect the four evolutionary sequences at the stages
when the
total mass of the star has been reduced to 2.9, 2.7, 2.5 and 2M$_{\odot}$.

In all four cases considered here
the  3M$_{\odot}$ models share a few common features, namely:
i) they experience HBB, which is usually associated
to an overluminosity\footnote{In quiescent stages
TP-AGB models with HBB are brighter than expected by the classical core-mass
luminosity relation (e.g.Boothroyd \& Sackmann 1991)} effect,
and ii) they enter the domain of C-stars,
as the surface C/O ratio increases above unity due to the
TDU. At the same time significant differences arise.

The S06H and S06C models experience a much weaker
mass loss at the beginning,
evolving at approximately constant mass for many TPs;
this is at odds with the behaviour of BH and BC models,
 where an efficient mass loss
determines an earlier extinction of HBB and its overluminosity
(following a rapid cooling of the envelope structure).
This circumstance is seen in the maximum luminosity attained,
which is $\sim 0.2$~dex fainter than in S06H and S06C sequences.

The role played by the opacity
treatment can be understood by examining the evolution
of models sharing the same description of mass loss, e.g.
the S06H and S06C models.
In the latter, the rapid increase in
the mass loss rate as soon as the surface C/O exceeds unity (clearly
detectable as a jump in $\dot M$) favours an earlier reduction of
the mass of the external mantle, which again causes an
earlier drop in the luminosity $L$ (we have a $\sim 0.1$~dex
difference in the maximum luminosity in this case).
The drop in $L$ is associated to a lower
temperature at the bottom of the convective envelope, i.e.
less favourable conditions for HBB.
As a consequence, we may conclude
that, in general, using a mass loss description only mildly dependent on the
luminosity, and/or neglecting the changes in the molecular chemistry
in the opacity computations when C/O$>1$,
correspond to larger temperatures
in the external mantle, in favour of a more efficient HBB.

\subsection{C-star stage and HBB quenching}
\label{ssec_nucleos}
Figure~\ref{mcore} shows the evolution of the C-O core mass during
the TP-AGB phase of models with different initial masses and/or
input proscriptions.
The BH and BC models (full squares and open triangles) evolve to smaller core
masses, as a result of the higher mass loss in their early TP-AGB phase.

\begin{figure*}
\caption{The same as Fig.~\ref{mcore}, but showing the temperature
at the bottom of the convective envelope
}
\label{tbce}
\end{figure*}

In the left panel we see that in the less massive BH and BC models
the mass of the remnant
is almost independent of the adopted opacity,
being only slightly higher, as expected, in the BH case. When
the \citet{oscar2} mass loss rate is used (S06H and S06C models) the
masses of the remnant differ more significantly
(see col.3 of Table \ref{yields}),
the discrepancy consisting in
$\delta M_{\rm c} \sim 0.08M_{\odot}$ for $M=2M_{\odot}$,
$\delta M_{\rm c} \sim 0.04M_{\odot}$ for $M=2.5M_{\odot}$, and
$\delta M_{\rm c} \sim 0.02M_{\odot}$ for $M=3.5M_{\odot}$.
The difference in $M_C$ is higher at lower masses, because these
models reach more easily the C-rich stage, that is
accompanied by an increase in the molecular opacity in the S06C models.
The stronger sensitivity of the S06H and S06C models to the opacity
treatment compared to the BH and BC models
can be explained by the fact the \citet{blo} formalism
predict quite high mass-loss rates so that the models become
C-rich when most of their envelope masses
have already been lost, with consequent little possibility
for establishing great differences in
the mass of the remnants.

The evolution of   $M_{\rm c}$  for higher mass models
is shown in the right panel of Fig.~\ref{mcore}.
The BH and BC lines are again almost over-imposed in this diagram,
and the differences between the S06H and
S06C sets of models persist up to masses of the order of $4M_{\odot}$,
whereas in more massive models the surface C/O ratio hardly approaches
unity, thus rendering the results almost opacity-independent. For the
$4.5M_{\odot}$ model we note that the tracks are practically split into
two branches, according to the mass loss treatment.

\begin{figure*}
\caption{Left: the variation of the surface carbon abundance in
models with initial mass 2M$_{\odot}$, calculated for various
treatments of mass loss and molecular opacities. Right: the same
as in the left panel, but for models with M=2.5M$_{\odot}$.
The meaning of the different tracks is as follows. solid: BH models;
dotted: BC; dashed: S06H; long-dashed: S06C
}
\label{carb1}
\end{figure*}

The description given above can be complemented with the
behaviour of the temperature at the bottom of the convective
envelope, $T_{\rm bce}$.
This is the key-quantity to understand the degree
of nucleosynthesis expected, and thus the chemical composition
of the ejecta.
The evolution of $T_{\rm bce}$ as a function of the current
stellar mass is shown for all models in Fig.~\ref{tbce}.

For $M\leq 2.5M_{\odot}$ HBB conditions, i.e. $T_{\rm bce} \ga
60\times 10^6$ K,  are never
reached, with the only exception of the S06H models,
where the relatively mild mass loss, due both to the
 \citet{oscar2} prescription itself and to the
molecular opacities that neglect CNO variations,
favours the increase in the core mass and temperature at the
bottom of the envelope, eventually reaching HBB conditions.

\begin{figure}
\label{carb2}
\end{figure}

This is confirmed by the two panels of Fig.~\ref{carb1},
showing the evolution of the surface carbon abundance
for the two models with initial masses $2.0$ (left panel) and $2.5~M_{\odot}$
(right panel).
In models BH (solid line) and BC (dotted line) the surface carbon
increases with a step-like profile that is typical of TDU effects.
The BC models undergo a smaller number of TPs as a consequence
of the higher mass loss rates experienced as the C/O ratio
exceeds unity, therefore the surface carbon is lower.

An increasing trend of the carbon abundance
is also shown by the S06C models (long-dashed),
although in this case most of the carbon enhancement is achieved
at the beginning, when the mass lost is negligible; at later stages
the strong mass loss triggered by the formation of C-bearing molecules
prevents further meaningful changes in the surface carbon mass
fraction.

Different is the behaviour of the
S06H models (dashed lines) in both panels of Fig.~\ref{carb1}:
 the fact that HBB is operating can be recognized from the
rapid drop of the surface carbon abundance (being converted into
nitrogen). In these cases the ejecta of the stars
are expected to be carbon-depleted.

Moving to higher masses, we see from Fig.~\ref{carb2} that in the
$3M_{\odot}$ model HBB conditions are reached in both the S06C and S06H
models, whereas in the models calculated with the \citet{blo} mass loss
HBB is quenched when the correct opacities are adopted (compare the dotted
and solid tracks in Fig.~\ref{carb2} and the
corresponding evolution of $T_{\rm bce}$ marked by open triangles and full 
squares in the bottom--left panel of Fig.~\ref{tbce}).
HBB is found in all models of higher mass, as can be seen in the
right panels of Fig.~\ref{tbce}. The only difference we see here is
the different mass left in the envelope when the asymptotic
temperature at the bottom of the envelope is reached. For the same
reasons previously discussed, S06H and S06C models reach this stage
when only a small amount of mass has been lost: this will have some
effects on the yields (see Sect.~\ref{sec_yields}).

\subsection{Summary}
The results  confirm the analysis by \citet{paola02}
and the main findings of the investigation by \citet{sergio}:
low-mass AGB models need to be calculated with the correct low-T
opacities, that account for the changes in the molecular chemistry
driven by variations of the C/O ratio, and more generally
in the CNO abundances.
Neglecting the surface carbon enrichment
in the opacity computations delays the ejection of the external
mantle, leads to a higher final core mass and, in more massive models,
leads to higher temperatures at the envelope base,
hence favouring  the development of HBB.

The extent of the changes introduced by a correct opacity computation
is conditioned to the treatment of mass loss. The indirect effects
of using C-rich opacities on the growth of the remnant core, maximum 
luminosity and temperature at the bottom of the convective envelope
are larger when the mass loss in the early AGB phase is sufficiently mild.
In fact, under these circumstances, the C-rich stage is reached when the
envelope still contains a large fraction of mass; this applies, for
instance, when a  treatment like the \citet{oscar2} is used.
Models calculated with the \citet{blo} formula are less sensitive to
the opacity changes.

The analysis on HBB made by \citet{paola07} is confirmed. For the
reasons mentioned above, the quenching of HBB in models calculated
with the \citet{blo} mass-loss description is restricted to a narrow
range of masses (clustering around 3M$_{\odot}$ in the present investigation).
Conversely, when a more moderate mass loss, e.g. the VW93 or the modification
suggested by \citet{oscar2} is used, the extinction of HBB involves a wider
range of masses, that in this investigation spans the interval
from $1.8$ to $2.6M_{\odot}$.

All models with M$\geq 3.5$M$_{\odot}$ achieve HBB conditions
regardless of the opacity treatment, because
the C/O ratio keeps in any case well below unity. Some differences
between models differing in the mass loss treatment persist up to
$\sim 4.5M_{\odot}$, and tend to disappear for larger masses. As
extensively discussed in \citet{paolo4}, in this range of masses the
key quantity, that controls the activation
and the strength of HBB, is the treatment of the convective instability,
and, specifically, the efficiency of convection.

\begin{figure}
\caption{The CNO variation in the ejecta for the 4 sets of
models under consideration.
The ordinate shows R(CNO), i.e. the ratio between the
average C+N+O in the ejecta and the initial C+N+O content of
the star. The meaning of the different lines is the same as in
Fig.~\ref{carb1}}
\label{cno}
\end{figure}

\section{How robust are the AGB yields?}
\label{sec_yields}
Changes in the surface chemical composition of AGB stars are caused
by HBB and TDU. The former keeps the overall C+N+O abundance
constant\footnote{We recall that the invariance
applies to the total C+N+O abundance
expressed as number fraction, and not as mass
fraction: in fact, the CNO-cycle preserves the total number of
the CNO catalysts, not their total mass.},
whereas TDU increases the carbon abundance, hence
the total C+N+O, independently of the possible later conversion
(via HBB) of $^{12}$C to $^{14}$N at the bottom of the convective zone.

\subsection{The C+N+O increase}
In the context of the distinctive chemical patterns exhibited by GC stars,
the surface C+N+O is a key-quantity in the debate concerning
the self-enrichment scenario by massive AGB stars \citep{annibale, paolo3},
because the spectroscopic surveys of GC stars confirmed that the
total C+N+O is approximately constant \citep{ivans}, even when
comparing stars with different surface abundances of other elemental species.
Therefore, AGB ejecta with a great CNO enhancement would rule out their role
as the possible cause of the observed patterns \citep{amanda06}.

The CNO increase in the ejecta is shown in Fig.~\ref{cno};
this is the same quantity reported in col.~10 of Tab.~\ref{yields}.
In all cases R(CNO) shows a maximum for M$\sim 2-2.5$M$_{\odot}$:
lower masses experience fewer TDUs, whereas in more massive models
the effects of TDU are reduced, due to dilution of the dredged-up,
C-rich material into a massive envelope.

For $M\ga 3.5M_{\odot}$,
similarly to other quantities discussed in Sect.~\ref{sec_mod}, R(CNO)
depends mainly on the mass-loss prescription;
the models split into two branches. S06H and S06C ejecta show a
greater enhancement in the total CNO, due to the larger number of TPs
and TDUs experienced. Interestingly,
for M$\sim 6$M$_{\odot}$ the two branches tend to
converge to unity (i.e. invariance of the total C+N+O), since these
models are characterised by quite efficient HBB and large luminosities,
which lead to a quick ejection of the envelope.

For $M < 3.5M_{\odot}$ the treatment of the molecular
opacities becomes more important: the CNO enhancement is lower for
S06C and BC models compared to S06H and BH.
The largest R(CNO) is reached by the S06H models (dashed line)
due to the great number of TPs experienced, while the others
(BH, BC and VW models) show more similar trends.

\begin{figure*}
\caption{Carbon and nitrogen (left panel),
and  oxygen and sodium (right panel) contents in the AGB ejecta.
Lines connect models with the same initial mass.
}
\label{cnona}
\end{figure*}

The left panel of Fig.~\ref{cnona} shows the chemical content of the
ejecta, on the [C/Fe] vs. [N/Fe] plane. Models with the same
mass are connected with continuous lines. The points in the lower-right
corner correspond to low-mass models ($M\leq 2.5M_{\odot}$) not 
experiencing any HBB, and are characterized by large [C/Fe] as a 
consequence of the TDU. In general, a large value
of [N/Fe] is a signature of HBB. The spread in [N/Fe]
relative to models with masses in the range 2--3M$_{\odot}$ is
a mere consequence of the different (if any) degree of HBB achieved
in these masses. S06H models (full triangles) constitute a
sort of upper envelope for the N enrichment: as discussed previously, these
models experience a large number of TDUs, and HBB is always efficient.

Note the position of the 1.5M$_{\odot}$ BC model,
indicated by the full square with the lowest [N/Fe]: the
chemical content of the mass ejected shows only a modest
increase in carbon and no change in nitrogen, because the
higher mass loss quickly leads to end of the AGB phase.
This is also confirmed in Fig.~\ref{cnona} (right panel)
by the position of the same point in the
[O/Fe]-[Na/Fe] plane,
corresponding to [O/Fe]=+0.4 and [Na/Fe]=0, i.e.
the initial chemical composition assumed for these stars.

Higher mass models experience a more powerful HBB, thus
exhibit a lower [C/Fe] in their ejecta. The relatively
modest [N/Fe] is due to the
lower efficiency of TDU in increasing the surface abundance of
carbon when the dredged-up material is diluted into a massive envelope:
for the same reason this limits the increase of nitrogen, which is
produced by proton-captures on carbon nuclei.

\subsection{Sodium}
The right panel of Fig.~\ref{cnona} shows the oxygen and sodium
content of the yields, in terms of [Na/Fe] vs. [O/Fe].
These two elements are the main targets
in the spectroscopic surveys of GCs, and the oxygen-sodium
anti-correlation is a well-established pattern observed in
practically all the GCs \citep{carretta}.

The ejecta of the lowest masses are O-enriched (we recall
that the initial abundance of our $\alpha-$enhanced mixture
is [O/Fe]=+0.4), due to TDU. The S06C models (open circles)
show a slightly higher [O/Fe], because they experience many TDUs,
and HBB is far from being activated, thus keeping the surface oxygen high.
The greater is the mass, the lower is the oxygen content of
the ejecta, due to HBB which destroys $^{16}$O via proton captures
in favour of $^14$N.

The behaviour of sodium at the surface of AGB stars is described in
details in \citet{paolo08} (see their figure 4).
The surface sodium content
is unchanged in the lowest masses (see also col.~8 in Tab.~\ref{yields}),
while increases in models (M$\geq 2$M$_{\odot}$) where the temperature
at the bottom of the envelope is high enough to convert  the dredged-up
$^{22}$Ne to Na . Similarly to nitrogen,
the largest differences in [Na/Fe] due to mass-loss and opacity
prescriptions are found in
those models with mild HBB (note again the
higher values attained by the S06H models). For higher temperatures
the destruction channel of Na via proton capture prevails over production,
leading to the lower [Na/Fe] for larger stellar masses.

\subsection{An overview on the uncertainties in the yields by AGB stars}
The analysis developed so far allows us to have a first
appraisal of
the uncertainties affecting the yields from AGB stars produced by two
main factors:  mass loss  and low-T radiative opacities.
We have explored two treatments of mass loss which, especially
in the early phases of the AGB evolution, differ considerably:
in this way our investigation should sample the present status
of indetermination in the chemical yields.

In agreement with previous investigations, we confirm that using
opacities coupled to the actual surface composition
is mandatory for masses below a threshold value, corresponding to the full 
activation of HBB, which in the present work corresponds to
$M~\sim~3.5~M_{\odot}$.
When the changes in the molecular chemistry are correctly accounted for
in the opacities,
the attainment of the C-rich stage (C/O$>1$)  is usually
accompanied by a sudden increase in the mass loss, which
favours the ejection of the whole mantle, speeding up the end
of the  AGB evolution.
The enrichment in carbon and CNO is consequently lower.

Typically, the changes in the yields,
determined by use of the correct opacities, are
more significant in models where a milder  dependency of mass loss on
luminosity is adopted. In fact, this would correspond
to a longer TP-AGB lifetime during which
the surface composition may be affected by i) a larger number
of TDU episodes, and ii) more favourable conditions for the
development of HBB.
Conversely, assuming a very efficient mass-loss formalism,
as predicted by a \citet{blo}-like law,
the accuracy loss introduced by unsuitable opacities
(e.g. valid for scaled-solar mixture)
is less dramatic, since the carbon-rich
stage is reached when a considerable fraction
of the mass of the envelope is already lost.

Within this general trend, the sensitiveness of the yields to the
model assumptions vary according to the mass of the star, as
outlined below.

The largest uncertainties in the yields belong to stars
for which the temperature at the bottom of the convective
envelope is just sufficient to ignite HBB, i.e. $1.8 - 3.0~M_{\odot}$.
In particular, [N/Fe] and [Na/Fe] may vary by $\sim~2$~dex,
depending on the assumed input prescriptions.
The oxygen yield is more robust.

At lower masses, i.e. $M~\sim~1.5M_{\odot}$, the results
are more affected by the mass-loss description: the TDU does not even take
place when the   \citet{blo} prescription is followed, being prevented
by a rapid expulsion of the external envelope.

Above $3.5~M_{\odot}$ the opacity treatment
become less influential, and most of the uncertainty in the yields
should be ascribed to the
the mass-loss treatment.

This investigation has also shown that, although the yields
may vary according to the adopted mass-loss rate in any range of mass,
the predicted trends between different elemental species do not depend
on the model assumptions, e.g. the slopes in both
[C/Fe]-[N/Fe] and [O/Fe]-[Na/Fe] diagrams are
preserved (see Fig.~\ref{cnona}).

\begin{figure}
\caption{Mass-loss rates experienced by a 2M$_{\odot}$
model, computed with the ATON code,
according to various mass-loss prescriptions available in the literature.
}
\label{conf}
\end{figure}

\section{A comparison with previous investigations}
\label{sec_compar}
In view of testing the reliability of AGB models,
we compare our results with investigations by other
research groups, focusing on the
the recent investigations by \citet{achim} and \citet{sergio2},
in which low-T opacities are suitably calculated by considering the
increase in the surface carbon abundance, an essential requirement
to correctly describe C-rich stage.

We consider, in particular, the models computed with same initial
mass, $2.0~M_{\odot}$, that include our
S06C and BC models,  the $Z=0.001$ model
by \citet{sergio2} (see their Tab.~3), and the $Z=5\times 10^{-4}$ model
by \citet{achim} (see their Tab.~B.4). Although the
initial chemical compositions
are not the same, (\citet{sergio2} use a scaled-solar mixture
with $Z=0.001$,
whereas \citet{achim} adopt an $\alpha-$enhanced mixture with
Z=0.0005), the main differences can be explained in terms
of the different input prescriptions.

One striking feature is the different
number of TPs experienced, that is $19$ in \citet{sergio2}, $21$ and $16$ in our
two S06C and BC cases, and only $6$ in \citet{achim}, despite of the
fact that this latter model corresponds to the lowest metallicity.
The small number of TPs
found in the \citet{achim} work is accompanied by a shorter AGB
life-time, which is $\approx 1/3$ of our estimated values.
Due to the absence of HBB, we rule out convection as a possible cause
of such a discrepancy, and focus on the other two, uncertain, physical inputs,
i.e. mass loss and the treatment of the convective borders,
particularly during the TDU phase.

Let us first focus on the effect produced by different mass-loss
treatments.
Figure.~\ref{conf} shows the evolution of the mass-loss rate in our
$2.0~M_{\odot}$ AGB model computed with the ATON code, assuming CNO-variable
molecular opacities and three choices for $\dot M$, namely:
\citet{oscar2} (open circles), \citet{blo} (open triangles), and
\citet{achim} (full squares).
We recall that this latter scheme assumes a Reimers' formula
up to a threshold pulsational period of $400$ days,
followed by the \citet{wachter08} rate when the C-star stage is reached.

We note that the \citet{oscar2} treatment favours rates significantly smaller,
particularly at the beginning of the AGB phase,
with differences up to $\sim~1.5$~dex compared to the others.
A smaller difference exists between
the \citet{blo} and the \citet{achim} recipes, the latter being higher
by $\sim~0.2-0.3$~dex during most of the evolution.

In the early AGB the \citet{achim} treatment predicts rates of the order
of $10^{-7}-10^{-6}~M_{\odot}$/yr, which are, however, not sufficient to 
diminish the total number of TPs so as to recover their original result ($6$).
This suggests that the main reasons for the smaller number of TPs in the 
\citet{achim} investigation compared to our models is an earlier transition to 
the C-star stage. This difference is due to two reasons: a) The modelling 
of convective overshoot from the bottom of the envelope (see their equation 
2), based on an exponential decay of the diffusive coefficient, with the same 
e-folding distance used for main-sequences width fitting. This approach determines 
a much deeper extension of TDU when compared with our models 
(where no overshoot is assumed), and with those by \citet{sergio2}, who adopt 
an exponential decay of the convective velocity from the inner border of the external mantle, with an 
e-folding distance tuned in order to allow the synthesis of a given amount of 
$^{13}$C available for neutron-captures. This amount of extra-mixing is also the 
plausible reason for the smaller number of TPs experienced by the \citet{sergio2}
model compared to our S06C, and to the different final core-masses ($0.67~M_{\odot}$, to 
be compared to $0.70~M_{\odot}$); b) the convective overshoot from the He-burning
shell, adopted by \citet{achim}, and neglected here and in \citet{sergio2}, which,
as described by \citet{falk3}, renders the pulses stronger. The models
presented here, though including no overshoot, are much more similar to those 
by \citet{sergio2} in terms of duration of the whole AGB phase and of the chemistry of 
the ejecta.

A detailed investigation of the TDU phenomenology, and the possible effects of assuming 
some convective overshoot from the bottom of the convective envelope, is 
beyond the scope of the present paper; the interested reader may find in \citet{falk}, 
\citet{falk2}, \citet{mowlavi} and \citet{oscar} a full discussion on the key issues 
relevant to address this problem.

Anyhow, the present study clearly illustrates that the AGB evolution is the product 
of a complex interplay between various processes: the treatment of the extra-mixing, 
for instance, has a remarkable impact not only on the chemical but also on the physical 
properties of these stars.

\section{Conclusions}
In view of improving the robustness of the results provided by AGB
modelling (or at least to quantify the uncertainty range),
we have investigated the sensitivity
of the main physical and chemical properties to the choices inherent the
input macro- and micro-physics.

We find a threshold mass, that is $M\simeq 3.5~M_{\odot}$ for
$Z=0.001$ in the present analysis,
(note that it also depends on the assumed overshooting from the central
regions during the core H- and He-burning), above which the results are
little sensitive to the opacity treatment, because HBB prevents
the surface C/O ratio to exceed unity.

In this range of masses mass loss
modelling plays a crucial role.
Models calculated with a treatment of the mass loss
only mildly dependent on the luminosity, such as the classic \citet{VW93}
or the \citet{oscar2} recipe used here, predict smaller rates
during most of the AGB phase, thus more TPs are experienced. This
also favours the growth of the luminosity and temperature at the bottom
of the convective zone, and a more advanced HBB nucleosynthesis: the yields
will contain the signature of proton-capture nucleosynthesis, e.g.
lower carbon and oxygen abundances, and higher nitrogen content.
However, the general patterns of the C-N and O-Na relations, and the very
small increase in the overall CNO abundance, are almost
independent of the opacity and mass-loss treatment: we expect convection
modelling plays a relevant role here, in agreement with the analysis
by \citet{paolo4}.

For masses smaller than $3.5~M_{\odot}$ the interplay between opacity and
mass-loss treatment is more tricky, because the lack of HBB favours the
increase in surface carbon. Generally speaking, we confirm the results
by \citet{paola02, paola07}, \citet{sergio, sergio2} and \citet{achim}: the use of the correct opacities
in the low-T regime speeds-up the loss of the stellar envelope,
via the strong winds associated to the general expansion of the
structure when the surface C/O ratio exceeds unity.
We expect in this case a smaller number of TPs, and a less advanced
nucleosynthesis (if any) at the bottom of the convective envelope.

Models calculated with a mass-loss rate steeply dependent on the luminosity
show a more homogeneous behaviour, whereas when a milder dependency of
$\dot M$ with luminosity, e.g. a relationship that relates $\dot M$ to
the stellar period, is adopted, many physical properties, e.g. core-mass,
temperature at the bottom of the convective zone, total number of thermal
pulses experienced, follow a completely different behaviour.
In this case the quenching of HBB determined by the use of the C-rich
opacities, predicted by \citet{paola07}, is confirmed, being a common
feature of all the models with $1.8-3.0~M_{\odot}$. A narrower range of
masses (around $\sim~3~M_{\odot}$) is otherwise affected by this uncertainty
when an efficient mass-loss prescription, like the \citet{blo}
formula, is adopted.

Contrary to the models experiencing HBB, the yields from lower mass
stars are highly uncertain and model-dependent: the average [C/Fe] in the
ejecta shows up an overall uncertainty of $0.7$~dex, though this difference
reduces to $\sim 0.3-0.4$~dex if models calculated with the same opacities
are compared. The situation for nitrogen and sodium is more extreme, as
they are particularly sensitive to the ignition of HBB. The differences
in [N/Fe] and [Na/Fe] can reach $\sim 2$~dex, and even when comparing 
models with the same assumptions for the opacity treatment, discrepancies 
of one order of magnitude still persist. The oxygen yields appear, instead, 
more stable.

These trends should be considered representative of AGB models
with a low metal content ($Z \sim  0.001$)
while they need to be further explored at larger metallicities
i.e. including those characteristic of the LMC
(Ventura et al., in preparation).
Moreover, in the present unresolved scenario about the efficiency of
TDU and mass loss, a valuable contribution may come from
synthetic AGB models, i.e. via
a thorough calibration of AGB properties as a function of
stellar mass and metallicity  based on a large set of
observables (HR diagrams including the MID infrared,
C/M star ratios, chemistry of atmospheres and planetary nebulae,
initial-final mass relation, integrated colors and brightness fluctuations).
This work is in progress.

\section*{Acknowledgments}
P.M. acknowledges financial support by the University of Padova
(60A02-2949/09), INAF/PRIN07 (CRA 1.06.10.03),
and MIUR/PRIN07 (prot. 20075TP5K9).

\end{document}